\documentclass[twocolumn,showpacs,prd]{revtex4}
\usepackage{graphicx}
\usepackage{dcolumn}
\usepackage{bm}
\begin{document}

\title{A Quantum Phase Transition in the Cosmic Ray Energy Distribution}
\author{A. Widom and J. Swain}
\affiliation{Physics Department, Northeastern University, Boston MA USA}
\author{Y.N. Srivastava}
\affiliation{Physics Department, University of Perugia, Perugia IT}

\begin{abstract}
We here argue that the ``knee'' of the cosmic ray energy distribution 
at $E_c\approx 1 \ {\rm PeV}$ represents a second order phase 
transition of cosmic proportions. The discontinuity of the heat capacity 
per cosmic ray particle is given by $\Delta c=0.450196\ k_B$. However 
the idea of a deeper critical point singularity cannot be ruled out by 
present accuracy in neither theory nor experiment. The quantum phase 
transition consists of cosmic rays dominated by bosons for the low 
temperature phase $E<E_c$ and dominated by fermions for high temperature 
phase $E > E_c$. The low temperature phase arises from those nuclei 
described by the usual and conventional collective boson models of 
nuclear physics. The high temperature phase is dominated by protons. 
The transition energy $E_c$ may be estimated in terms of the  
photo-disintegration of nuclei.
\end{abstract}


\maketitle

\section{Introduction \label{intro}}

We have recently\cite{Widom:2014} discussed the power law energy 
exponent values \begin{math} \{\alpha \} \end{math} in the cosmic ray 
particle flux distribution\cite{Gaisser:1990,Stanev:2002} 
\begin{equation}
\left[\frac{d^4\bar{N}}{dt dA d\Omega dE}\right] \approx 
\frac{1.8\ {\rm Nucleons}}{\rm sec\ cm^2\ sr\ GeV}
\left(\frac{1\ {\rm GeV}}{E}\right)^\alpha .
\label{intro1}
\end{equation}
The more detailed measured exponent \begin{math} \alpha \end{math} 
in reality takes on two values depending upon the cosmic ray particle 
energy. For energies lower than the crossing energy 
\begin{math} E_c\approx 1 \ {\rm PeV}  \end{math}, the 
exponent takes on a boson value of 2.701178. For energies higher 
than the crossing energy the exponent takes on the fermion value
of 3.151374. The crossing energy is, of course, the location of 
the so called ``knee'' in the energy distribution.

In that the energy distribution depends on the heat and thereby 
entropy of evaporation of bosons from the cosmic ray source, 
there exists a quantum phase transition of cosmic proportions at 
the crossing energy. It is evidently a quantum phase transition 
since the order parameter involves the difference between 
Bose and Fermi statistical phases. In virtue of the experimental 
continuity of the energy distribution and thereby the entropy, 
the phase transition is higher than first order. To the 
theoretical and experimental present accuracy, the phase 
transition is of second order with discontinuities in the second 
derivatives of the entropy although more complicated non-analytic 
singularities cannot be ruled out. In Sec.\ref{igt}, the 
thermodynamic properties of ultra-relativistic ideal gases are 
reviewed. The heat capacity discontinuity is described 
in Sec.\ref{hc}.

The boson phase arises from those evaporation nuclei 
described by the conventional collective {\em Boson models} of 
nuclear physics\cite{Ring:2004}. There exist pairing correlations 
in {\em odd odd} nuclei made up of deuterons. Correlations 
between two spin one deuterons lead to spin zero alpha 
particles and so forth all within the pairing condensate. The 
condensate resides near the surface of evaporating high baryon 
number \begin{math} A\gg 1 \end{math} nuclei. For example, a neutron star 
itself is merely a nucleus of extremely high baryon number 
(\begin{math} A\gg \cdots \gg 1  \end{math}) with 
superfluidity (and superconductivity) in the neighborhood 
of the nuclear surface\cite{future}.

The ``partons'' from the pairing condensate are evidently the 
nucleons. The partons then turn into fermions for the phase that 
exists above the crossing energy. For energies above the knee, 
the cosmic rays must be composed mainly of protons. In order to 
comprehend the phase transition, it is necessary to understand 
how photons can photo-disintegrate the compound boson
{\em odd odd} nuclei ultimately into its nucleon 
parts\cite{Feynman:1998}. In Sec.\ref{rk} we employ a simple 
physical kinetics model of photo-disintegration to estimate 
the crossing energy. There is satisfactory agreement with 
experiment. A general view of the quantum cosmic ray phase 
transition is given in the concluding Sec.\ref{conc}.

\section{Ideal Gas Thermodynamics \label{igt}}

In the ultra-relativistic limit wherein the single particle 
energies are much larger than either the rest mass energy 
and the source environmental chemical potentials, the 
energy \begin{math} {\cal E} \end{math}, 
pressure \begin{math} P \end{math} and volume 
\begin{math} {\cal V} \end{math} of the gas are related by 
\begin{equation}
{\cal E}=3P{\cal V}.
\label{igt1}
\end{equation} 
Eq.(\ref{igt1}) holds strictly true for ``massless'' gas at 
``zero'' chemical potential. For such a case, the equation of 
state for the ideal gas reads 
\begin{equation}
P{\cal V}={\cal N}k_B \vartheta \ \ \ 
{\rm wherein}\ \ \ \vartheta =\left(\frac{\alpha }{3}\right)T,
\label{igt2}
\end{equation} 
wherein  
\begin{eqnarray}
\alpha_{MB}=3
\ \  ({\rm Maxwell-Boltzmann\ Statistics}), 
\nonumber \\ 
\alpha_{BE}\approx 2.701178 
\ \ ({\rm Bose-Einstein\ Statistics}), 
\nonumber \\ 
\alpha_{FD}\approx 3.151374 \ \ ({\rm Fermi-Dirac\ Statistics}),
\label{igt3}
\end{eqnarray} 
i.e. 
\begin{eqnarray}
\nonumber \\ 
({\rm Maxwell-Boltzmann\ Statistics})\Rightarrow \ \ \ \ 
\nonumber \\
P{\cal V}={\cal N}k_BT, 
\nonumber \\ 
({\rm Bose-Einstein\ Statistics})\Rightarrow \ \ \ \ 
\nonumber \\ 
P{\cal V}\approx 0.9003926\ {\cal N}k_BT, 
\nonumber \\ 
({\rm Fermi-Dirac\ Statistics})\Rightarrow \ \ \ \ 
\nonumber \\ 
P{\cal V}\approx 1.050458\ {\cal N}k_BT,
\label{igt4}
\end{eqnarray} 
wherein \begin{math} {\cal N} \end{math} is the number of 
particles. One may verify the Bose-Einstein case by 
computing the pressure of a photon gas, i.e. black body 
radiation. One may verify the Fermi-Dirac case by computing 
the pressure of a gas of Weyl neutrinos. 

The pressure from the Bose-Einstein gas is slightly 
lower than that of a Maxwell Boltzmann gas in that the quantum 
statistics describes an attraction. The pressure from the 
Fermi-Dirac gas is slightly higher  
than that of a Maxwell Boltzmann gas in that the quantum 
statistics describes a Pauli exclusion repulsion. Although 
the value change in alpha due to quantum statistics is small,
\begin{equation}
\Delta \alpha = (\alpha_{FD}-\alpha_{BE})
\approx 0.450196 ,
\label{igt5}
\end{equation}
Eq.(\ref{igt5}) is entirely responsible for the quantum cosmic 
ray second order phase transition.

\subsection{Heat Capacity \label{hc}}
 
The heat capacity at constant volume obeys 
\begin{equation}
{\cal C}_{N,V}=
\left(\frac{\partial {\cal E}}{\partial T}\right)_{\cal V,N}.
\label{hc1}
\end{equation}
In virtue of Eqs.(\ref{igt1}) and (\ref{igt2}) in a regime in 
which \begin{math} \alpha  \end{math} is uniform, the heat 
capacity per particle 
\begin{math} c = ({\cal C}_{N,V}/{\cal N}) \end{math} is related 
to the energy per particle 
\begin{math} E = ({\cal E}/{\cal N}) \end{math}
\begin{equation}
E=cT=\alpha k_BT.
\label{hc2}
\end{equation}
The heat capacity discontinuity is thereby 
\begin{equation}
\Delta c=k_B \Delta \alpha \approx 0.450196 \ k_B
\label{hc3}
\end{equation}
in virue of Eq.(\ref{igt5}).

\subsection{Entropy \label{ent}}

If \begin{math} s(E) \end{math} represents the energy per 
particle, then \begin{math} (1/T)=(ds/dE) \end{math} determines 
the temperature. Eq.(\ref{hc2}) then turns into a differential 
equation; It is
\begin{equation}
E=\alpha k_B \left(\frac{dE}{ds}\right),
\label{ent1}
\end{equation}
with the solution 
\begin{equation}
s(E)=\alpha k_B \ln\left(\frac{E}{\tilde{\epsilon }}\right). 
\label{ent2}
\end{equation}
In a single phase regime, the entropy of evaporation determines 
the energy distribution\cite{Widom:2014} via 
\begin{equation}
e^{-s(E)/k_B}=
\left(\frac{\tilde{\epsilon }}{E}\right)^\alpha .
\label{ent3}
\end{equation}
On either side of the crossover energy, the distribution of 
energy is determined by Eq.(\ref{ent3}). 

\section{Photo-Disintegration \label{PD}}

The cross-over energy will here be estimated on the basis of the 
physical kinetics. From this point of view, one goes from the 
lower temperature boson phase to the higher temperature fermion 
phase via photo-disintegration processes.

\subsection{Relativistic Kinematics \label{rk}} 

Consider a process wherein an initial compound nucleus 
\begin{math} I \end{math} is hit by a photon 
\begin{math} \gamma \end{math} and disintegrates into final fragments 
\begin{math} F \end{math},  
\begin{equation}
\gamma +I \to F, 
\label{rk1}
\end{equation}
with the four momentum conservation 
\begin{equation}
\hbar k_\gamma +p_I = p_F. 
\label{rk2}
\end{equation}
The invariant mass squared \begin{math} s \end{math} of 
the reaction Eq.(\ref{rk1}) obeys 
\begin{equation}
c^2 s = -p_F^2=-(p_I + \hbar k_\gamma )^2 = 
- p_I^2 -2\hbar k_\gamma \cdot p_I 
\label{rk3}
\end{equation}
for a {\em real} (not virtual) photon 
\begin{math} k_\gamma ^2 =0 \end{math}. Thus 
\begin{equation}
c^2(M_F^2-M_I^2) = -2\hbar k_\gamma \cdot p_I . 
\label{rk4}
\end{equation}
In the rest frame of the initial compound nucleus, the 
photon frequency \begin{math} \varpi_\gamma \end{math} 
obeys 
\begin{math} -k_\gamma \cdot p_I =M_I\varpi_\gamma \end{math}
so that 
\begin{equation}
\hbar \varpi_\gamma = 
(M_F-M_I)c^2\left[\frac{M_F+M_I}{2M_I}\right], 
\label{rk5}
\end{equation}
wherein the term in square brackets on the right hand side 
of Eq.(\ref{rk5}) takes into account the total recoil of the 
final state fragments. 

In the rest frame of the cosmic ray 
source, the photon frequency is given by 
\begin{eqnarray}
\omega_\gamma (1-\cos \theta ) = \varpi_\gamma  
\left[\frac{c^2M_I}{E}\right], 
\nonumber \\ 
\hbar \omega_\gamma (1-\cos \theta )= 
c^4\left[\frac{M_F^2-M_I^2}{2E}\right],   
\label{rk6}
\end{eqnarray}
wherein \begin{math} E \end{math} is the energy of the initial 
compound nucleus and \begin{math} \theta \end{math} is the 
angle between the photon and compound nuclear three momenta.  
Averaging over angle and types of final fragmentation products 
yields \cite{GZK}
\begin{equation}
\overline{\hbar \omega_\gamma}=
c^4\overline{\left[\frac{M_F^2-M_I^2}{2E}\right]}. 
\label{rk7}
\end{equation}
The central Eq.(\ref{rk7}) of this section enables us to 
estimate the crossing energy. 

\subsection{Crossing Energy \label{ce}}

We have discussed above that  
\begin{math} 
\overline{\hbar \omega_\gamma }\approx 3k_BT_\gamma 
\end{math}, 
wherein \begin{math} T_\gamma \end{math} the radiation 
temperature of the source. The crossing energy can 
thereby be estimated as 
\begin{equation}
E_c \approx c^4
\left[\frac{\overline{M_F^2-M_I^2}}{6k_BT_\gamma}\right]. 
\label{rk8}
\end{equation}
The orders of magnitude involved in the estimate 
Eq.(\ref{rk8}) are 
\begin{equation}
E_c \sim \frac{(100\ {\rm MeV})^2}{10\ {\rm eV}}
\sim 1\ {\rm PeV}. 
\label{rk9}
\end{equation}
The crossing energy in Eq.(\ref{rk9}) is in satisfactory 
agreement with experiment.

\section{Conclusion \label{conc}}

We have argued that the well observed {\em knee} in the cosmic 
ray energy spectrum corresponds to a quantum phase transition 
from a lower temperature boson dominated cosmic ray beam to a 
higher temperature fermion dominated cosmic ray beam. The lower 
energy boson dominated regime owes its existence to the collective 
boson nuclei at the sources. These lead to the quantum symmetry 
energy contribution to the semi-empirical nuclear mass 
formula  \cite{Schuck:2004}. The higher energy fermion phase owes 
its existence to the free single nucleons which are the partons 
from which the bosonic nuclei were constructed. As the energy per 
particle increases there is a de-confinement quantum statistical 
phase transition as the bosons are disintegrated into  
constituent fermions.

There are many important applications of this picture of 
cosmic ray structure. Some light has been shed upon the structure of 
ordinary nuclei in terms of collective boson models. The 
notion of a superconducting shell near the surface of neutron 
stars has been explored. Clearly, nuclear transmutations 
and the cosmic ray sources of different chemical elements should 
now be studied anew. With regard to these future developments, 
we are at the beginnings. However, we hope that our studies of 
cosmic ray dynamics have clarified their deep connections to 
fundamental processes through rather precise determinations of 
the two correct critical indices along with an estimate of the 
cross-over energy .

\section{Acknowledgements}

J.S. thanks the National Science Foundation for its support via NSF grant PHY-1205845.


\begin{thebibliography}{14}
\bibitem{future} Condensates of bosonic nuclei can form
distinct superfluid and superconducting states other than
the traditional neutron-neutron and proton-proton pairings
usually considered, which we plan to discuss in detail in
a separate publication.
\end{thebibliography}
\end{document}